\documentclass[doublecol]{epl2} 

\usepackage{adjustbox}
\usepackage[numbers]{natbib}

\usepackage{longtable}
\usepackage{caption}
\usepackage{amsmath}

\title{Investigating the Bulk Properties of Charged Particles in Different $\eta$ Bins Using a Modified Tsallis Model}
\shorttitle{Properties of nuclear matter using Tsallis model} 

\author{H.I. Alrebdi\inst{1} \and Muhammad Ajaz\inst{2} \and Murad Badshah\inst{2} \and M. Waqas\inst{3} \and Nourah A.M. Alsaif\inst{1}}

\institute{                    
  \inst{1} Department of Physics, College of Science, Princess Nourah bint Abdulrahman University, P.O. Box 84428, Riyadh 11671, Saudi Arabia\\
  \inst{2} Department of Physics, Abdul Wali Khan University Mardan, Mardan 23200, Pakistan \\
  \inst{3} School of Mathematics, Physics and Optoelectronic Engineering, Hubei University of Automotive Technology 442002, Shiyan, People's Republic of China   
}

\abstract{
This paper presents a comprehensive analysis of the double-differential \(p_T\) distributions of charged particles in twelve distinct pseudorapidity regions of equal width in \(pp\) collisions at center-of-mass energies of 0.9, 2.36, and 7 TeV. Utilizing the modified Tsallis function with mean transverse flow velocity, our study demonstrates a very good agreement between experimental data and the model employed. The fit quality is consistently high across all \(p_T\) ranges, as assessed by Data/Fit panels accompanying each plot. Extracted parameters, including kinetic freeze-out temperature (\(T_0\)), transverse flow velocity (\(\beta_T\)), non-extensivity parameter (\(q\)) and mean transverse momentum $\langle p_T \rangle$ dependencies are shown on pseudorapidity (\(\eta\)) and collision energy (\(\sqrt{s}\)). \(T_0\), \(\beta_T\) and $\langle p_T \rangle$ exhibit a decreasing trend with increasing \(\eta\) due to lower in energy transfer along high $\eta$ regions, while they show a heightened sensitivity to \(\sqrt{s}\). \(q\) increases with \(\eta\), indicating a closer thermal equilibrium in mid-\(\eta\) particles. The paper also explores correlations among these parameters, emphasizing relationships between \(T_0\), \(\beta_T\), \(q\) and $\langle p_T \rangle$. Our study provides valuable insights into the thermal and dynamic characteristics of high-energy proton-proton collisions, contributing to the broader understanding of the bulk properties of nuclear matter produced in these interactions.
}

\begin{document}

\maketitle



\section{Introduction}
\label{sec:intro}
The measurement of global properties in collisions holds significant importance, particularly in the context of studying non-perturbative aspects of Quantum Chromodynamics (QCD) and in constraining phenomenological models and event generators. These interactions primarily involve soft processes or those with small-momentum transfer. Investigating such scenarios is vital for a thorough understanding of the conditions under which challenging and unconventional measures of interactions are observed.
An important way to characterize the overall features of high-energy proton-proton (pp) collisions is to analyze the transverse momentum and pseudorapidity distributions of charged particles produced in these collisions. Over a long length of time, this particular observation has been the focus of both large-scale experimental studies and theoretical investigations. The origins of producing particles at collider energy is closely related to the interaction of perturbative (hard) and non-perturbative (soft) QCD processes. Only phenomenological modeling techniques are able to adequately represent the major contributions of parton hadronization and soft scattering processes at lower transverse momenta. Thus, measurements inside this range provide useful constraints for improving models and event generators, especially with regard to hadron-collider and cosmic-ray physics.

In this present investigation, we conduct analyses utilizing the modified Tsallis function to examine the inclusive primary charged-hadron multiplicity densities, expressed per unit transverse momentum ($p_T$) and pseudorapidity ($\eta$), within the pseudorapidity range $|\eta| < 2.4$, divided into twelve equal intervals at 0.9, 2.36 \cite{source1}, and 7 TeV \cite{source2}. Here, $p_T$ represents the transverse momentum relative to the beam axis, and $N_{ch}$ signifies the count of charged hadrons within specific $\eta$ or $p_T$ intervals. The pseudorapidity is defined as $-\ln[\tan(\frac{\theta}{2})]$, where $\theta$ is the polar angle of the particle in relation to the anti-clockwise beam direction.
By definition, primary charged hadrons encompass all charged particles generated in interactions, including those originating from strong and electromagnetic decays, while excluding products of weak decays and hadrons resulting from secondary interactions. Particles with appropriate lifetimes smaller than 1 cm are included in the primary count of charged hadrons. Products that result from secondary interactions are not included in this count, though. To account for prompt leptons, an additional correction is done at a precent level.
Such measurements, along with their dependence on collision energy, play a pivotal role in comprehending the mechanisms of hadron production and the distinct contributions of soft and hard scattering in the LHC energy regime. Additionally, these measurements provide a reference point for studying nuclear-medium effects in PbPb collisions at the LHC.

Profound developments in the empirical detection of different particle types have prompted in-depth studies of the statistical properties of hadron spectra \cite{Tawfik:2014eba,Yassin:2019xxl}. Boltzmann-Gibbs (BG) statistics were initially used extensively to characterize particle multiplicities, their fluctuations, and correlations \cite{Magalinskii1957,Fast:1963uql,Fast:1963dyc}. The ultimate stages of high-energy multiparticle manufacturing processes were not adequately taken into consideration by this method \cite{Zalewski:2004fz}. An extensive substitute was made available with the advent of Tsallis statistics \cite{Tsallis:1987eu}. The standard exponential and logarithmic functions are introduced with \(q\)-counterparts in Tsallis statistics, in contrast to BG statistics, where \(q \neq 1\). A nonextensive framework for particle creation is provided by this expansion of the statistical space \cite{Yamano2002}. When the system enters a non-equilibrium state due to high correlations and possible nontrivial ergodicity, BG statistics are less significant and the Tsallis technique becomes more important \cite{Tawfik:2004sw}.
The \( p_T \) distributions of final-stage hadrons in proton-proton collisions at the RHIC and LHC experiments have been well-modeled by a number of Tsallis function modifications \cite{Cleymans_2012, CLEYMANS2013351, PhysRevD.83.052004, olimov}. For correct \( q \) value extraction, it is important to investigate extended \( p_T \) ranges, since the Tsallis function's non-extensivity parameter \( q \) is especially sensitive to the high \( p_T \) region (\( p_T > 3 \) GeV/c) \cite{PhysRevD.95.056021, universe5050122}.

Different models have been combined with Tsallis statistics to include transverse flow extraction from the \( p_T \) distributions of high-energy heavy-ion and proton-proton collisions at the RHIC and LHC. The Blast-Wave model with Tsallis statistics (TBW model) \cite{PhysRevC.79.051901, liu}, and the improved Tsallis distribution with transverse flow effect \cite{bhatt, doi:10.1142/S0217732320501151} are the notable one while the Blast-Wave model with Boltzmann–Gibbs statistics (BGBW model) \cite{PhysRevC.81.024911, PhysRevC.79.034909}, and the Hagedorn function with transverse flow \cite{universe5050122, Khandai_2014, doi:10.1142/S0217732320502375, doi:10.1142/S0217751X21501499} are among the frequently utilized models to estimate the transverse expansion velocity and kinetic freeze-out temperature.
The goal of this publication is to compare nonextensive statistical analyses of all charged particles (defined above) spectra as a function of pseudorapidity at different center of mass energies. The structure of the paper is as follows: The formalism is described in Section \ref{sec:formalism}, where the nonextensive nonequilibrium Tsallis statistics are introduced. Results are presented and discussed in Section \ref{sec:RsltDsc}. Section \ref{sec:conc} provides an overview of the conclusions.

\section{The Data, Formalism, and Method} 
\label{sec:formalism}
The study in Ref. \cite{source1, source2} presents measurements of the distributions of $p_T$ and pseudorapidity for inclusive charged hadrons in non-single-diffractive (NSD) proton-proton collisions at centre-of-mass energies of 0.9, 2.36 and 7 TeV. Being the sum of non-diffractive and double-diffractive the NSD proton-proton collisions yield charged hadrons in the full range of $\eta$. These measurements have been conducted using data collected by the Compact Muon Solenoid (CMS) detector at Large Hadron Collider (LHC). The data have been collected in twelve different $\eta$ intervals. The charged hadron yield has been derived through three distinct methods: (i) tallying reconstructed clusters within the pixel barrel detector, (ii) assembling pixel track pairs from clusters in various pixel barrel layers, and (iii) reconstructing tracks across the entire tracker volume by combining pixel and strip hits.

The simplest form of the Tsallis distribution function used to describe the $p_T$ spectra of the particles produced in high energy collisions is given as \cite{zheng2016comparing},
\begin{equation}
 \frac{1}{N_{ev}} \frac{1}{2\pi p_T} \frac{d^2N}{dydp_T} = C'\biggl(1 + (q - 1) \frac{m_T}{T_{eff}}\biggr)^{-q/(q-1)} \label{Eq. 1}  
\end{equation}
Here, the transverse mass \(m_T\) is defined as the square root of the sum of the square of the transverse momentum \(p_T\) and the square of the rest mass \(m_0\) of the observed particle. The parameter \(T_{eff}\) (Effective Temperature) is not a real temperature. it contains information about the effects of both average thermal motion in the form of kinetic freeze-out temperature ($T_0$), and flow effects in the form of transverse flow velocity ($\beta_T$). The parameter \(q\) serves as the non-extensivity parameter, giving deviations from the standard Boltzmann-Gibbs distribution. A value of \(q\) nearing 1 suggests a distribution resembling the conventional exponential distribution, whereas deviations from \(q = 1\) signal non-extensive effects, possibly arising from particle interactions or correlations. The symbol \(C'\) represents the Normalization Constant, ensuring that the probability density integrates to unity across the entire range of values. It serves to appropriately scale the distribution function to align with experimental data.

The Tsallis distribution given in Eq. (\ref{Eq. 1}) does not obey all thermodynamic laws and hence is a thermodynamically inconsistent one. The thermodynamically consistent Tsallis equation in its simple form being used to fit the experimental data is given by,
\begin{equation}
 \frac{1}{N_{ev}} \frac{1}{2\pi p_T} \frac{d^2N}{dydp_T} = Cm_T\biggl(1 + (q - 1) \frac{m_T}{T_{eff}}\biggr)^{-q/(q-1)} \label{Eq. 2}  
\end{equation}
Here, the newly appeared symbol \(C\) represents the Normalization Constant which is different from $C$ but has the same role as discussed for $C'$ above.

In comparison to the works in Ref. \cite{olimov, olimov2021multiplicity, Khandai_2014} where the two effects i.e., $T_0$ and $\beta_T$ are disentangled by introducing the embedded transverse flow velocity in the conventional Hagedorn function, the present study aims to introduce the similar flow effects into the thermodynamically consistent Tsallis distribution given in Eq. (\ref{Eq. 2}). The effects were introduced into Eq. (\ref{Eq. 2}) for the first time in Ref. \cite{k.olimov} where it was utilized to describe the mid-y $p_T$ distributions of the identified charged particles in different centrality intervals of Xe + Xe collisions at $\sqrt{s_{NN}}$ = 5.44 TeV. The modified form of the Tsallis equation is given by the following equation where $m_T$ is replaced with $<m_T>(m_T-p_T<\beta_T>)$ to introduce the transverse flow velocity \cite{k.olimov},
\begin{equation}
\begin{split}
\frac{1}{N_{ev}} \frac{1}{2\pi p_T} \frac{d^2N}{dydp_T} = C <\gamma_T>\Bigl(m_T-p_T<\beta_T>\Bigr) \times \\
\biggl(1+<\gamma_T>\frac{(q-1)(m_T-p_T<\beta_T>)}{T_0}\biggr)^{-q/(q-1)} \label{Eq. 3} 
\end{split}
\end{equation}
The function utilizes the following parameters: \(<\gamma_T>\) calculated as \(<\gamma_T>\) = $1/\sqrt{1-\beta_T^2}$ is the average boost factor, where \(<\beta_T>\) represents the average flow velocity, indicating the collective speed of particle flow; \(T_0\) stands for the kinetic freezeout Temperature, indicating the temperature at which particles cease interactions, thereby maintaining the transverse momentum \(p_T\) of the particle constant; and \(q\) denotes the non-extensivity parameter as defined earlier.
In the current work, we used Eq. (\ref{Eq. 3}) to fit the experimental data to study its bulk properties.

\section{Results and discussion}
\label{sec:RsltDsc}
Fig. 1 presents the double differential $p_T$ distributions as functions of transverse momentum ($p_T$) of different charged particles in various pseudorapidity ($\eta$) classes, produced in $pp$ collisions at centre-of-mass energies of 0.9, 2.36 and 7 TeV. The experimental data used in our analysis are sourced from references \cite{source1, source2}, and within our figures, these data points are displayed by various distinct geometric symbols.

\begin{figure*}[hbt!]
\begin{center}
\hskip-0.153cm
\includegraphics[width=5.5cm]{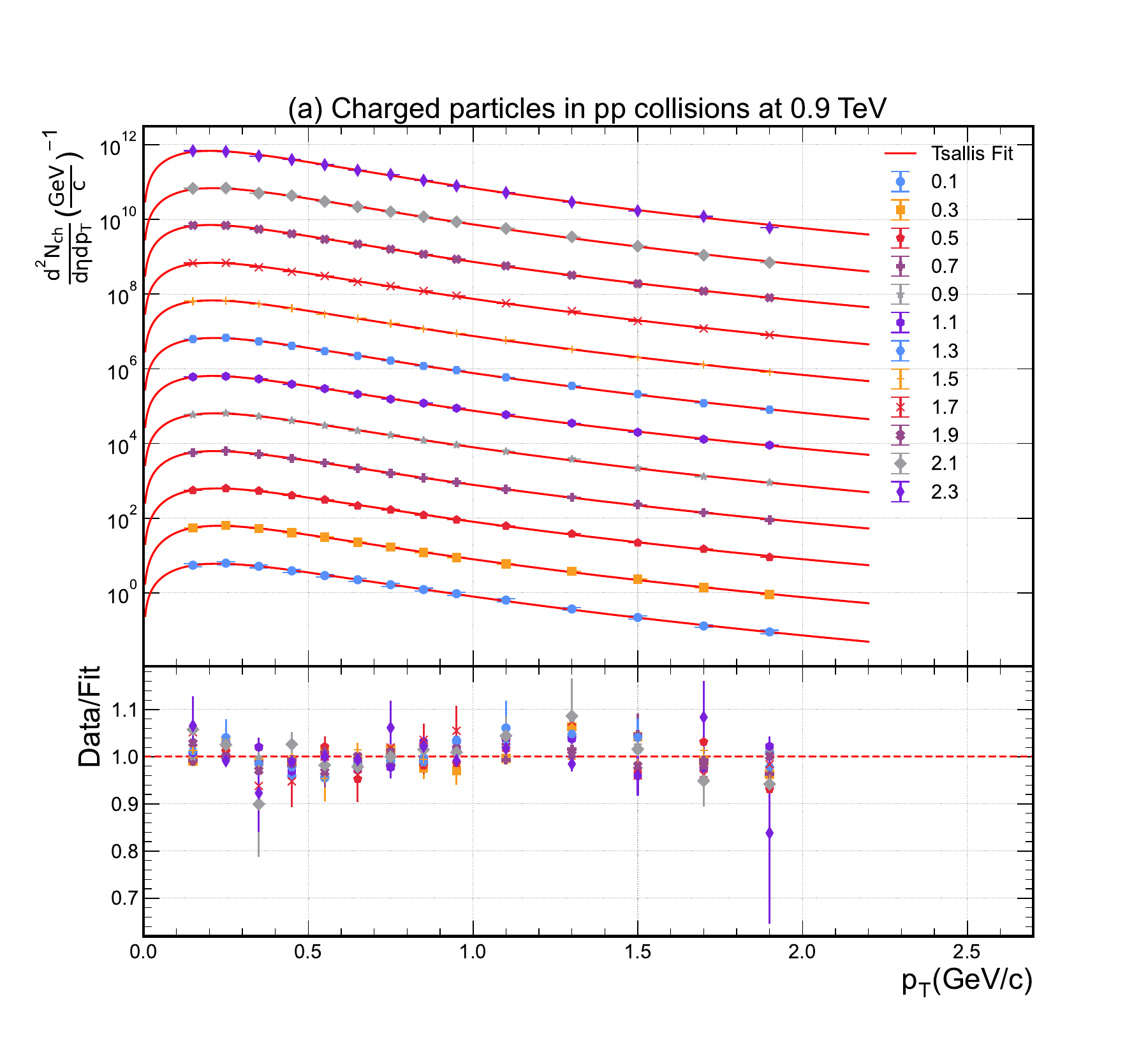}
\includegraphics[width=5.5cm]{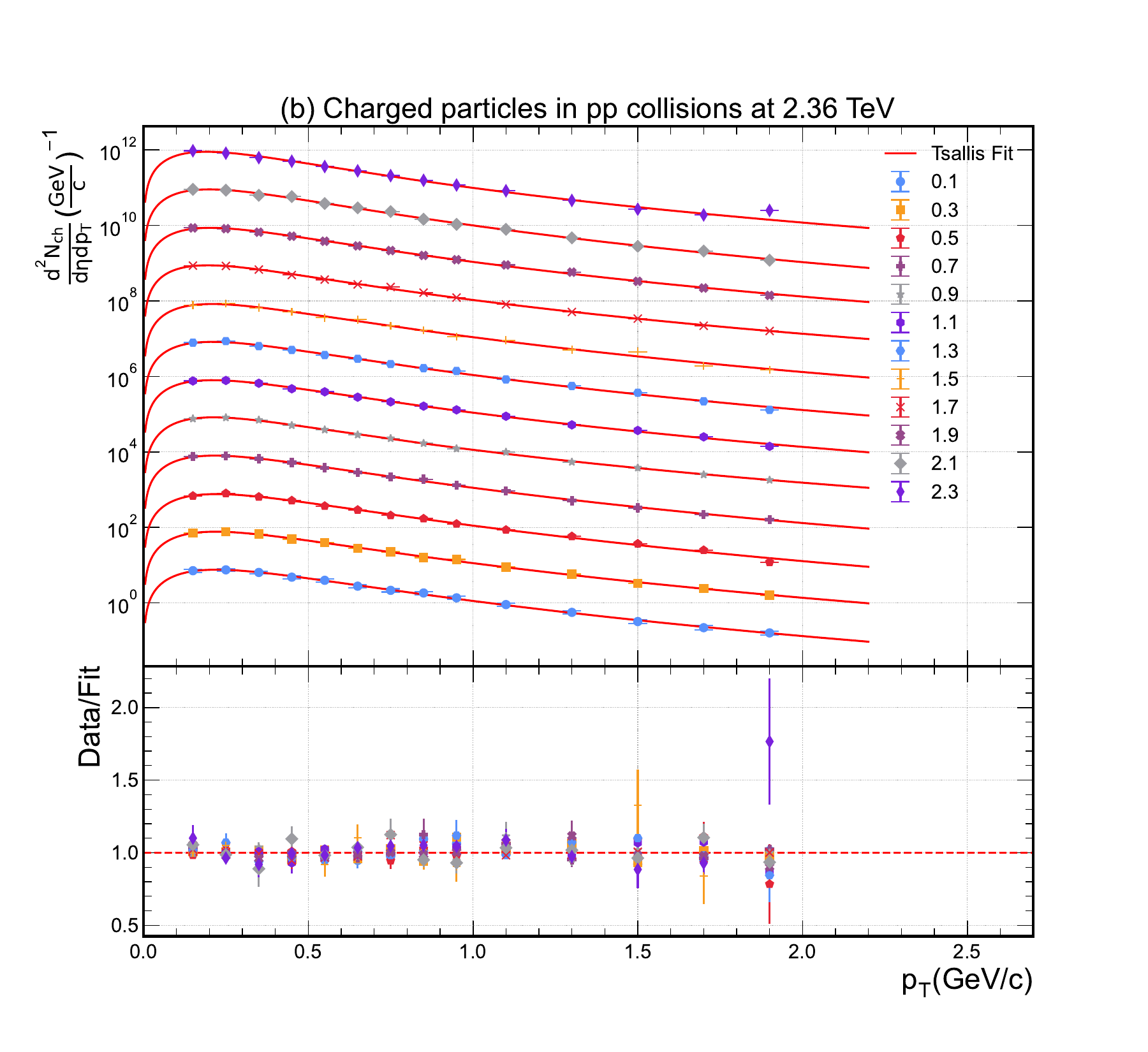}
\includegraphics[width=5.5cm]{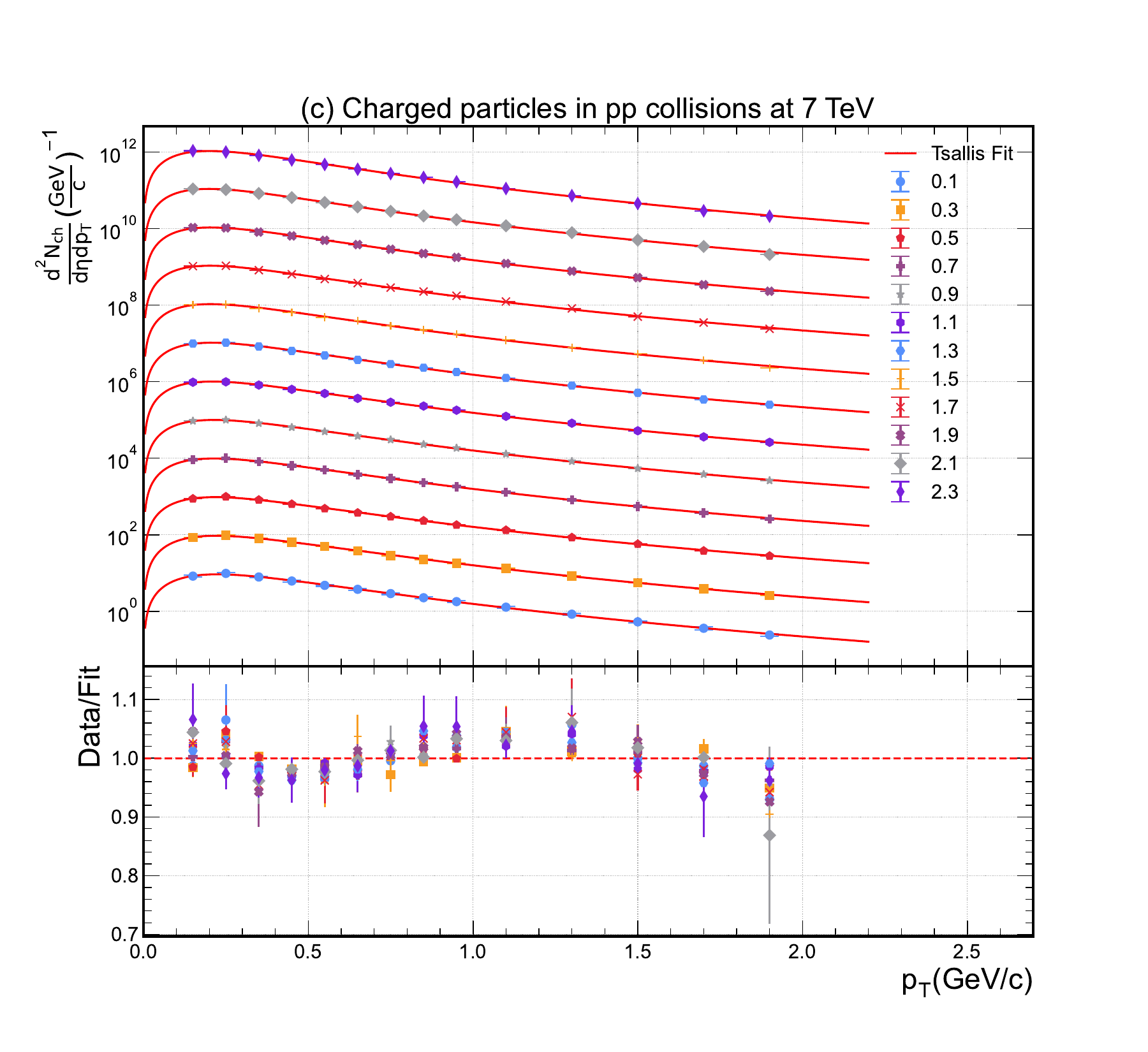}
\end{center}
\caption{The $p_T$ distribution of charged hadrons in various $\eta$ intervals produced in p--p collisions at 0,9, 2.36 and 7 TeV collision energies. The different coloured data points represent the experimental data for different $\eta$ values, while the solid lines are the fit results of Eq. (\ref{Eq. 3}). Each plot has a Data/Fit panel at its bottom which measures the fit quality of Eq. (\ref{Eq. 3}). \label{fig:1} }
\end{figure*}

\begin{figure*}[hbt!]
\begin{center}
\includegraphics[width=5.5cm]{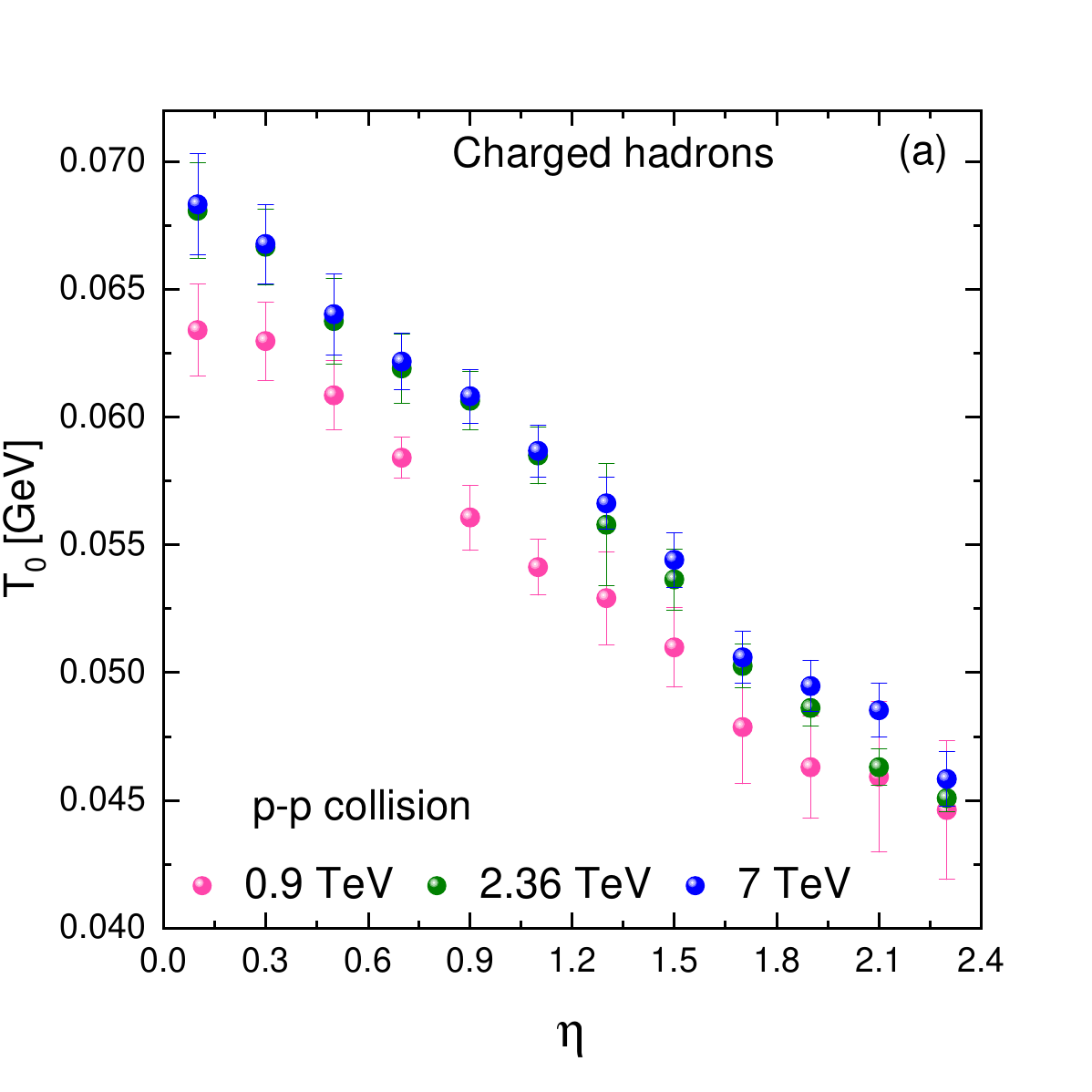}
\includegraphics[width=5.5cm]{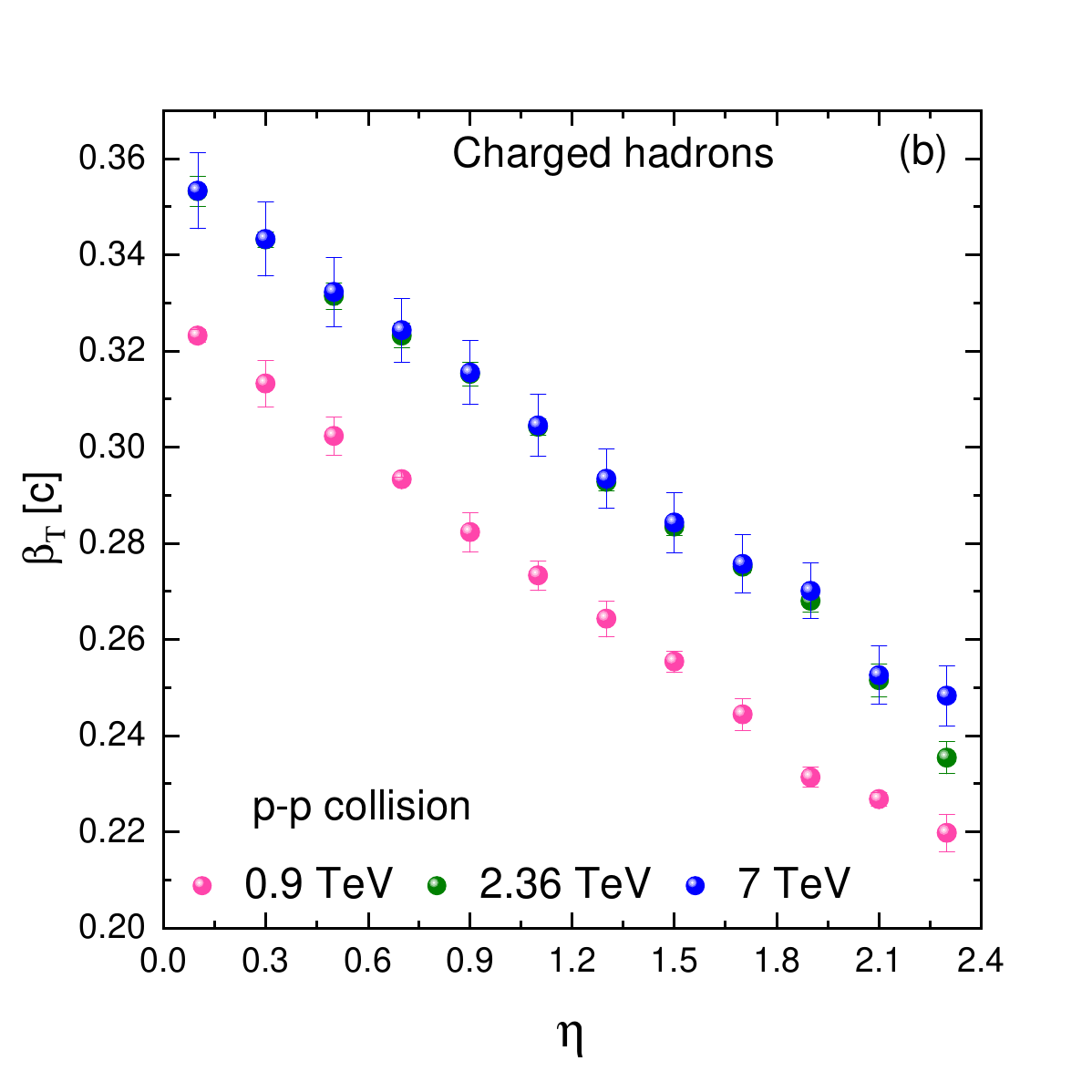}
\includegraphics[width=5.5cm]{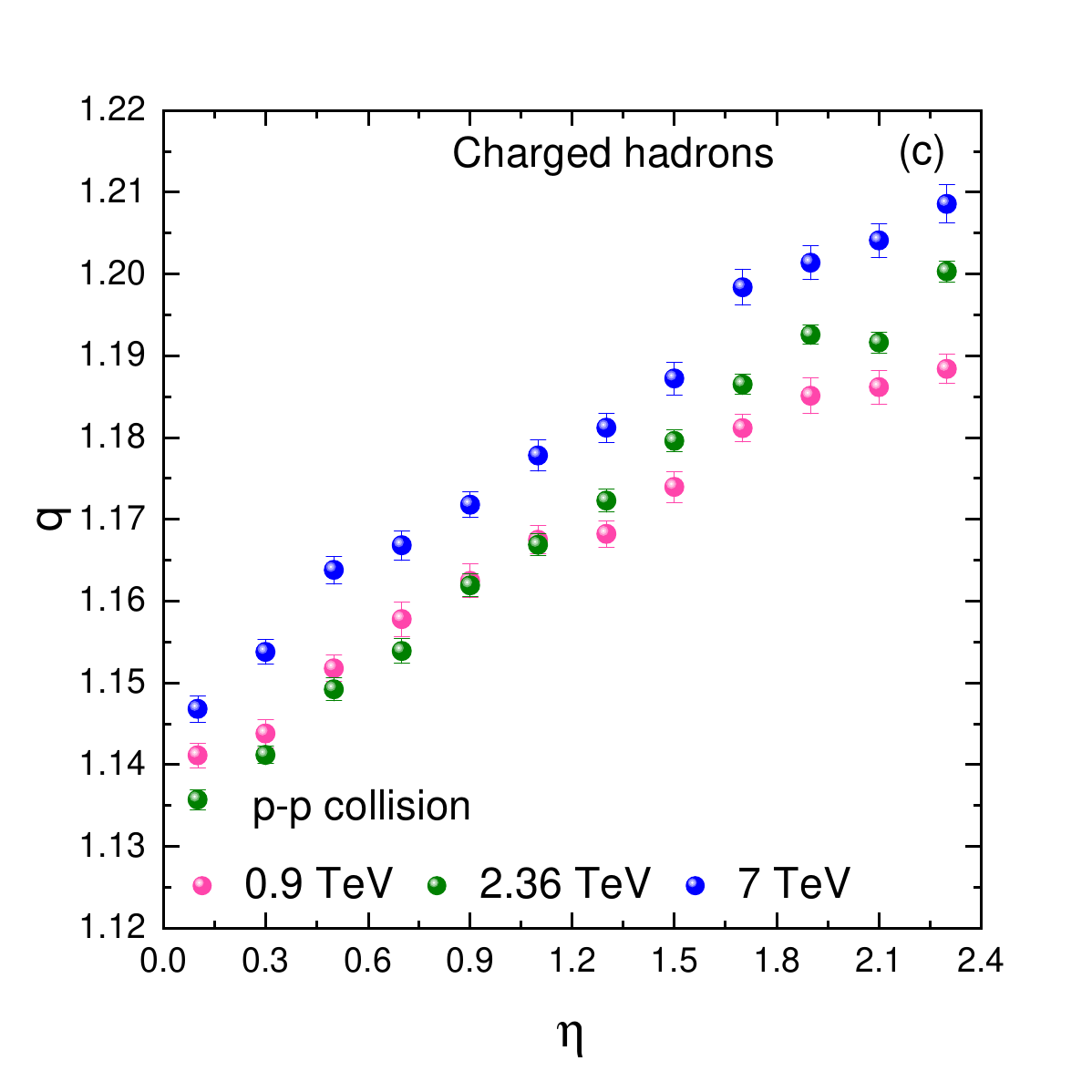}
\includegraphics[width=5.5cm]{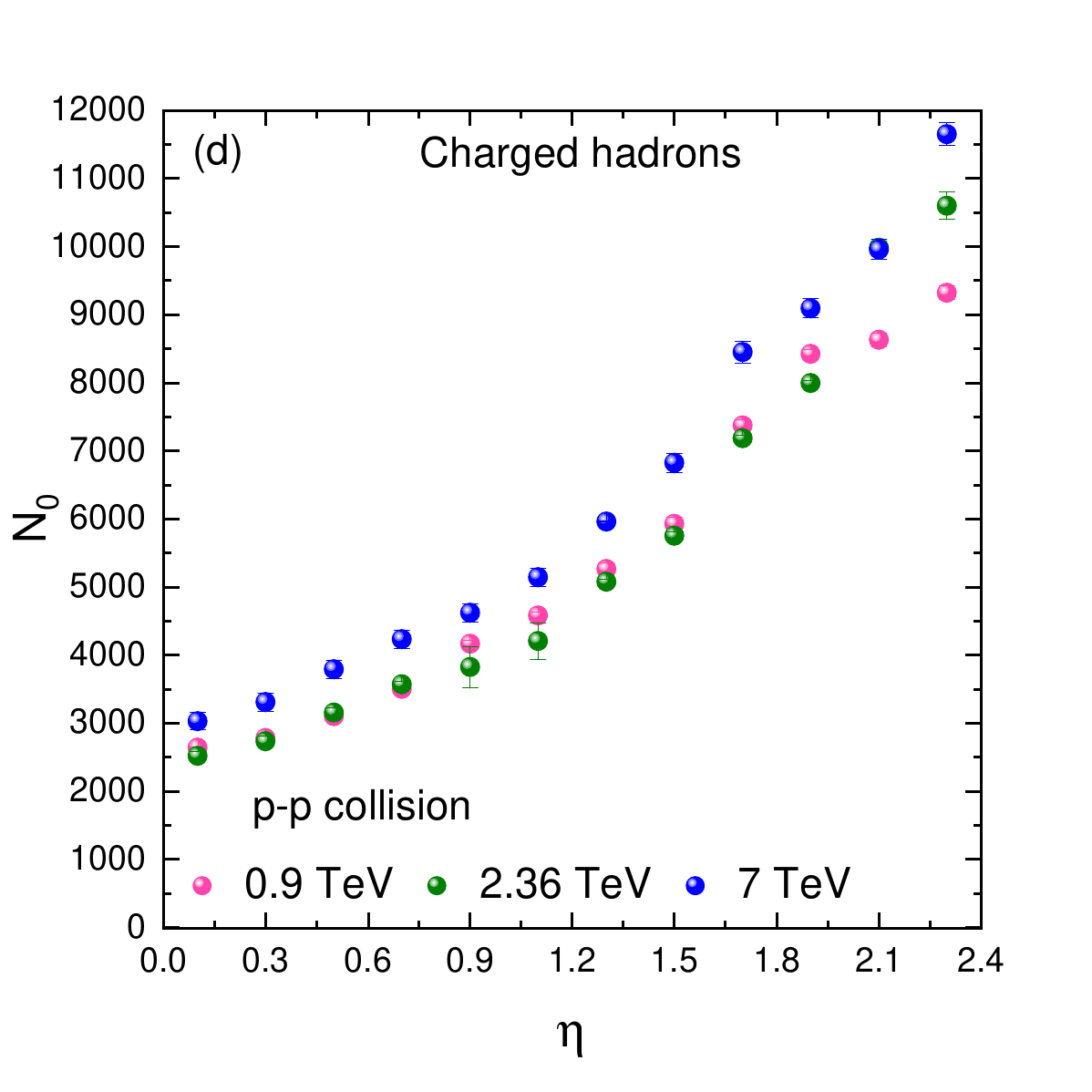}
\includegraphics[width=5.5cm]{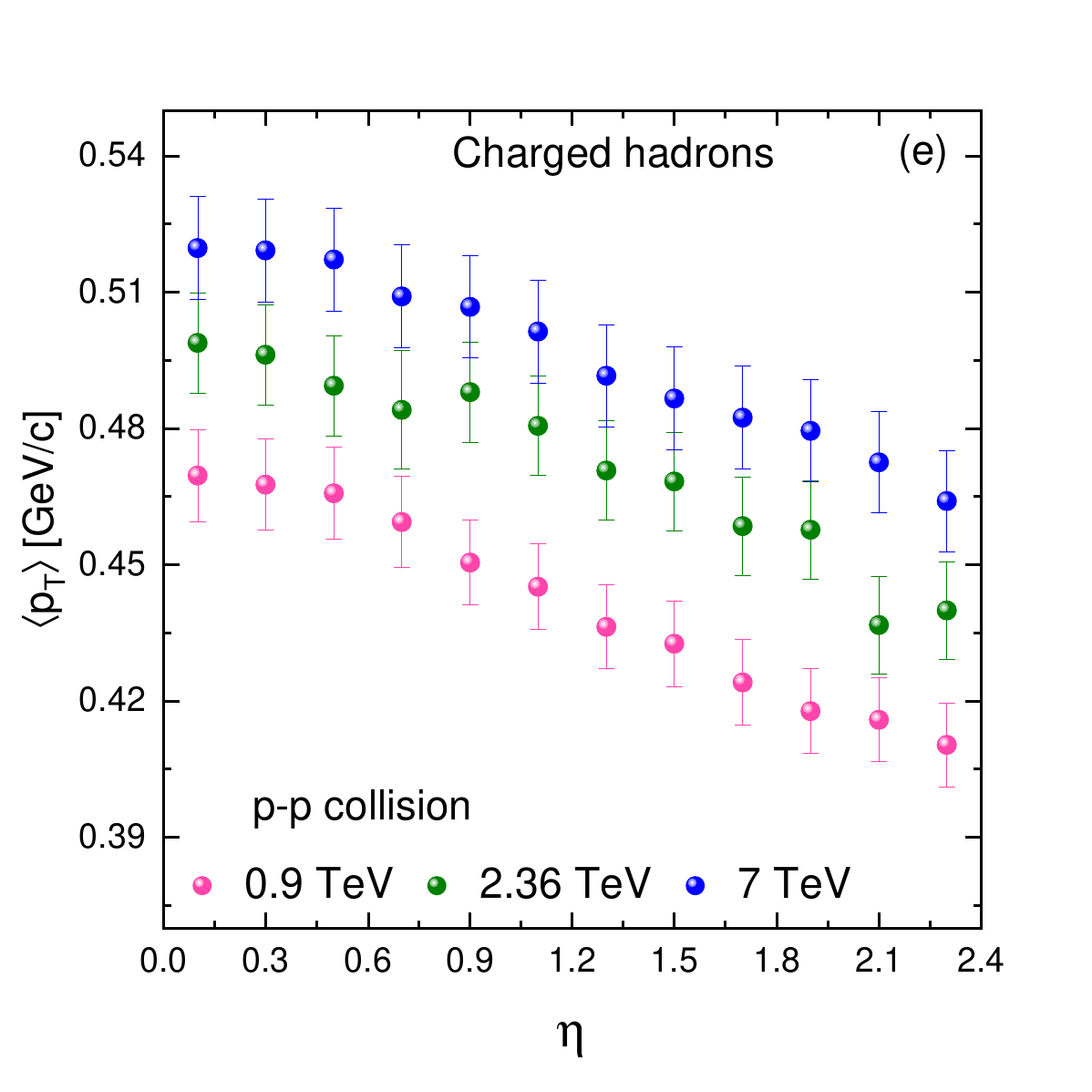}
\end{center}
\caption{(a) $T_0$, (b) $\beta_T$, (c) q, (d) fitting constant $N_0$ and (e) $\langle p_T \rangle$ as a function of $\eta$ in pp collision at 0.9, 2.36 and 7 TeV.\label{fig:2} }
\end{figure*}

\begin{figure*}[hbt!]
\begin{center}
\includegraphics[width=5.5cm]{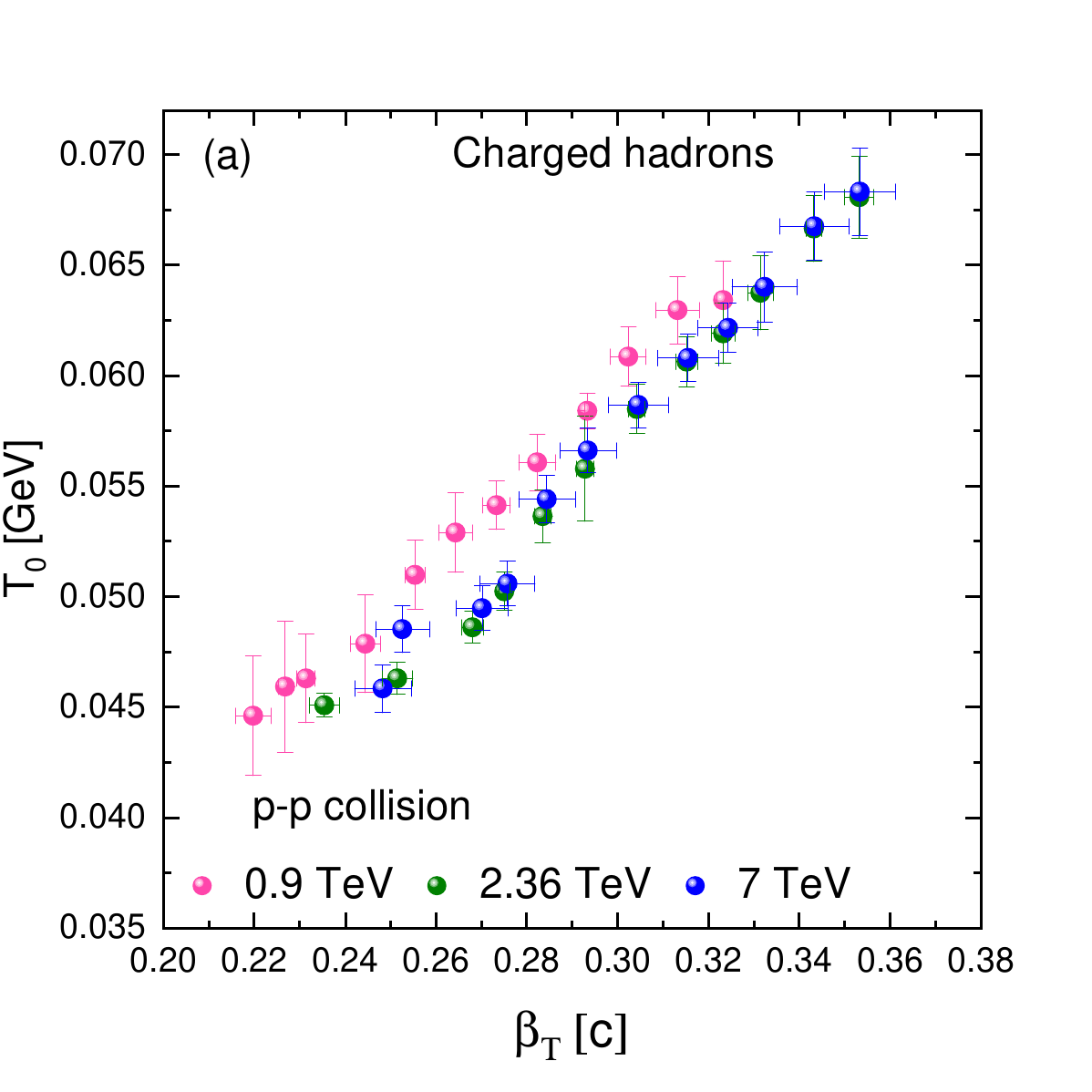}
\includegraphics[width=5.5cm]{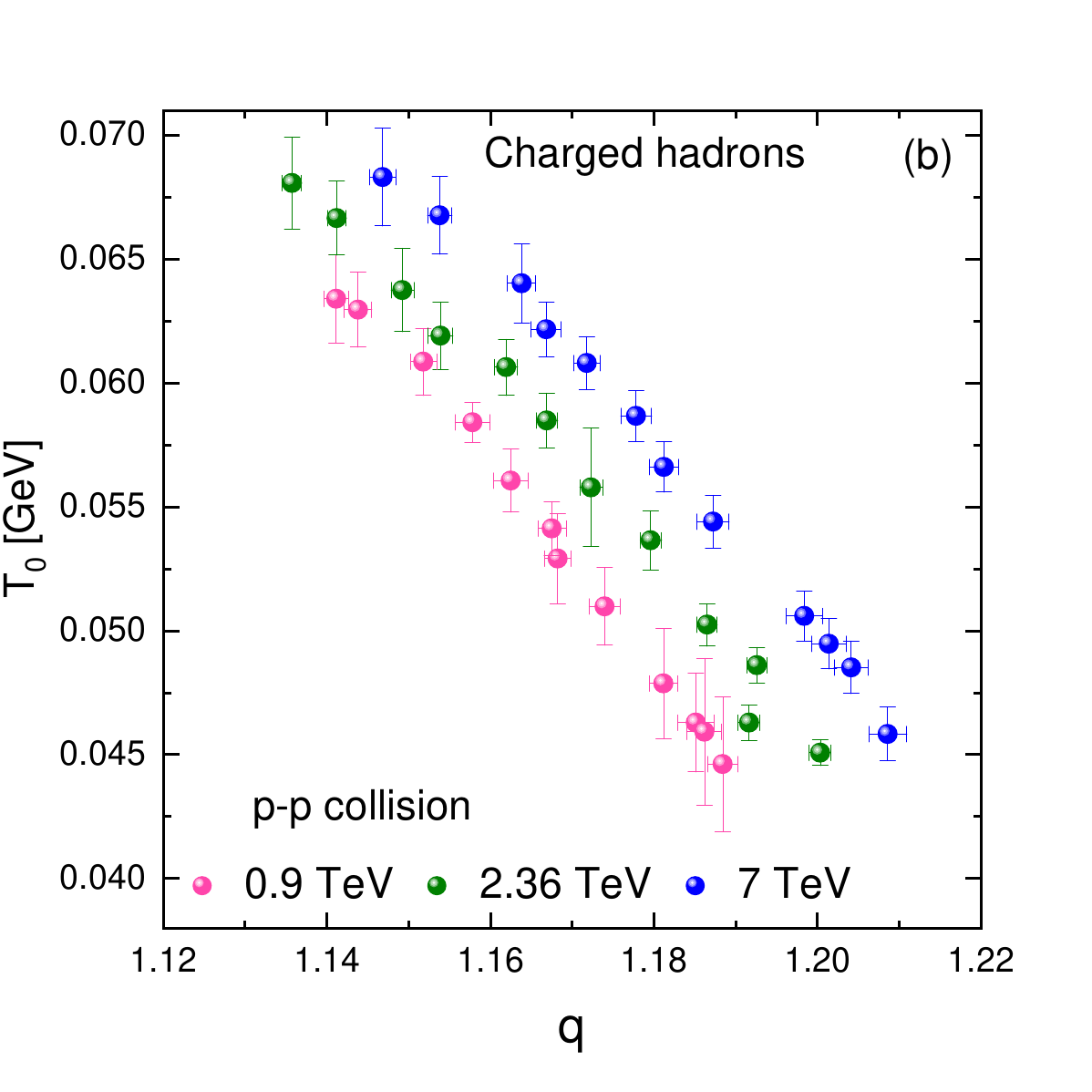}
\includegraphics[width=5.5cm]{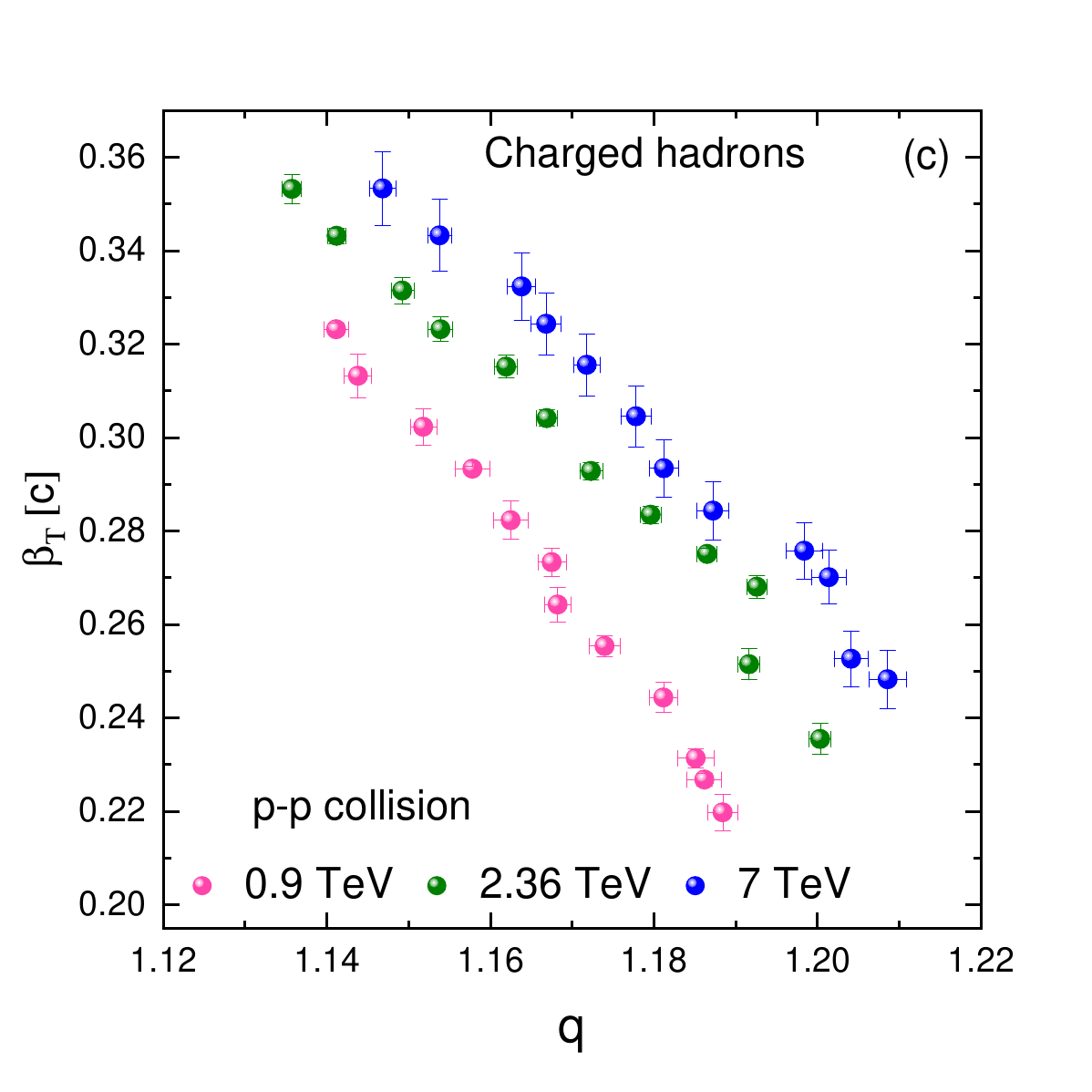}
\includegraphics[width=5.5cm]{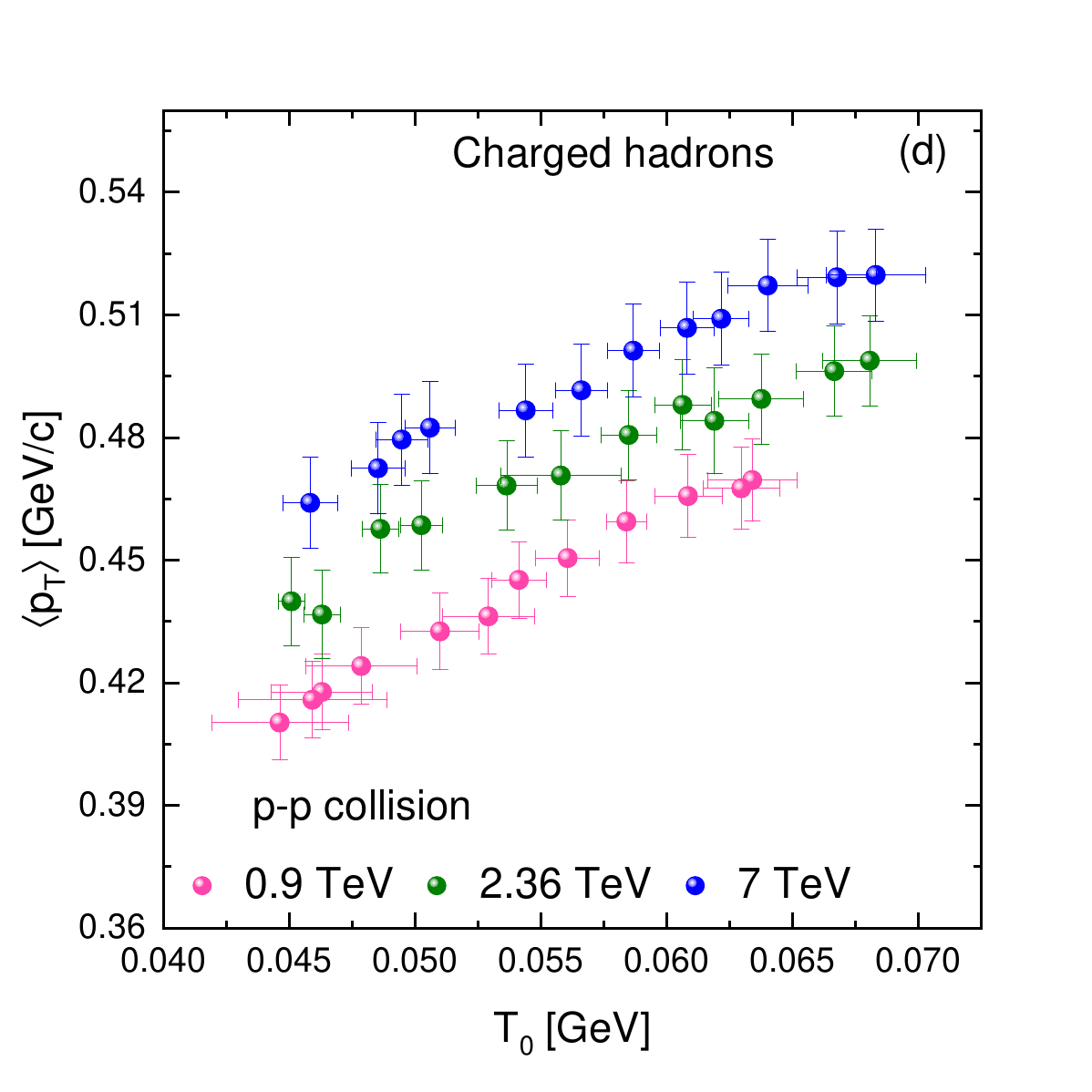}
\includegraphics[width=5.5cm]{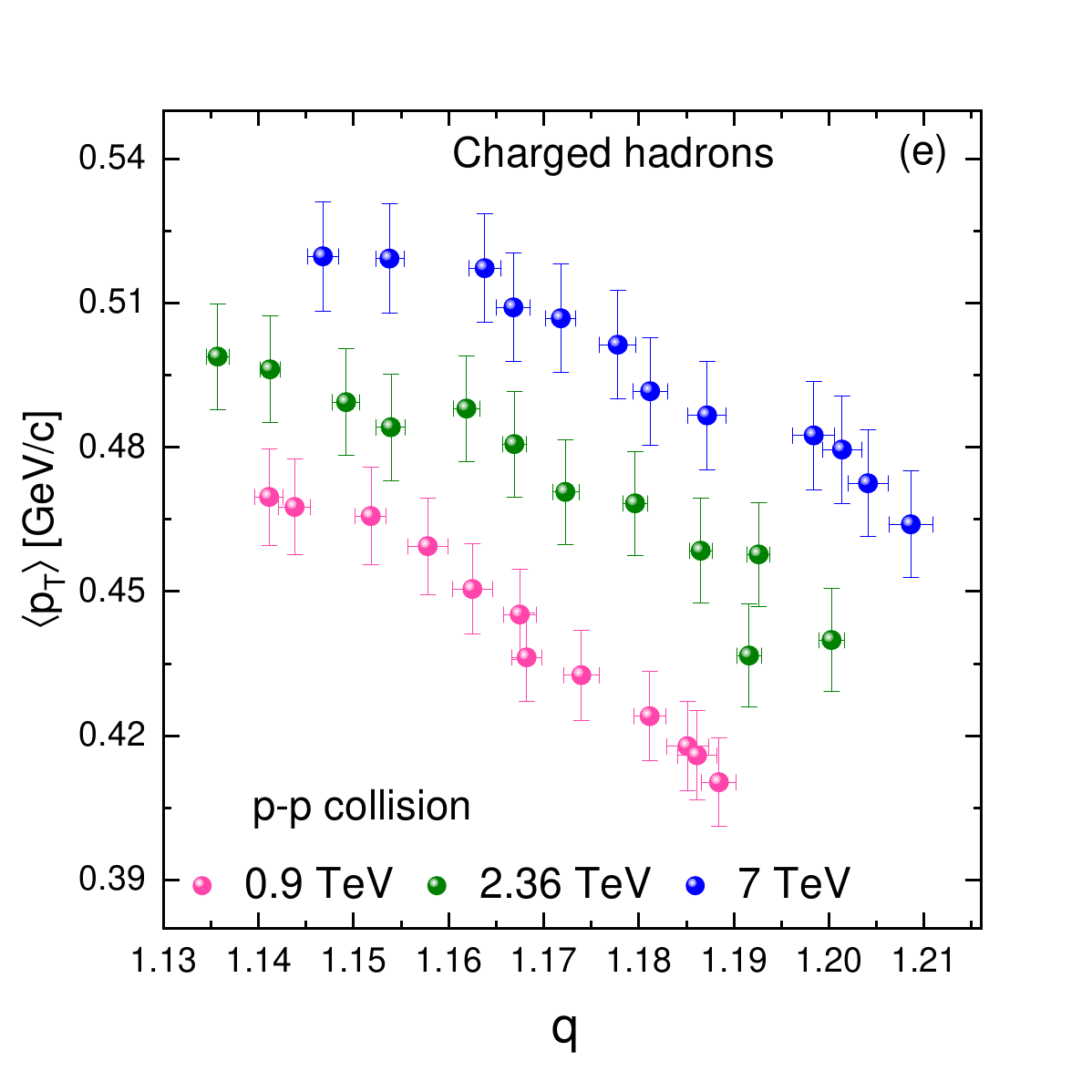}
\includegraphics[width=5.5cm]{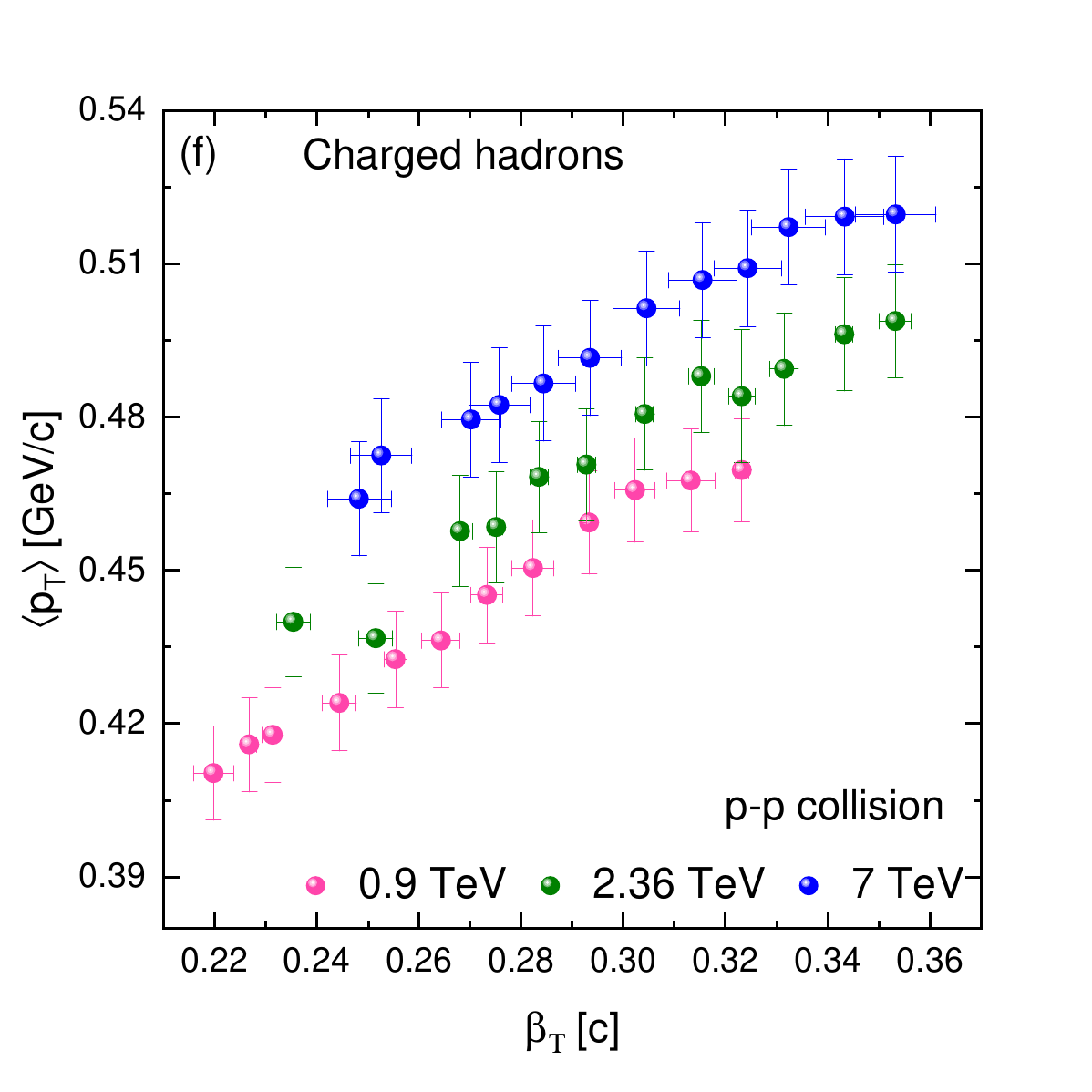}
\end{center}
\caption{(a) $T_0$ vs $\beta_T$, (b) $T_0$ vs q and (c)$ \beta_T$ vs q for charged hadrons produced in pp collision at 0.9, 2.36 and 7 TeV.\label{fig:3} }
\end{figure*}
In Fig. 1, solid lines are used to represent the results of the modified Tsallis function. It is clear from the figure that there is a good agreement between the experimental data and the model we used. This agreement holds across the entire range of $p_T$. To assess the quality of the fit, each plot within these figures is accompanied by a Data/Fit panel at its lower section. This panel provides a measure of the fit quality. Typically, a more promising fit is characterised by data points that are around the value of 1.

\begin{table*}[t!]
\begin{center}
\caption{The values of free parameters, the normalization constant, $\chi^2$ and $ndof$ obtained through the fitting procedure of the experimental data with modified thermodynamically consistent Tsallis distribution function given in Eq. (\ref{Eq. 3}).}
\begin{adjustbox}{width=\textwidth,center}
\begin{tabular}{ccccccccc}
\hline
\hline
Energy & $\eta$ & $T_{0}$ (GeV/$c$) & $q$ & $\beta_T$ &$\langle p_T \rangle$ & $N_0$ & $\chi^2$ & $ndof$   \\
\hline
    & 0.1 & 0.06341$\pm$0.01786  & 1.14110$\pm$0.00150 & 0.32320$\pm$0.02616 & 0.4696$\pm$0.0102 & 2641.75508$\pm$115.8066  & 3.03651  &  11  \\
    & 0.3  & 0.06297$\pm$0.01522  & 1.14380$\pm$0.00170 & 0.31320$\pm$0.09503 & 0.4676$\pm$0.0101 & 2781.18218$\pm$112.1516  & 2.25321  &  11  \\
    & 0.5 & 0.06086$\pm$0.01348  & 1.15180$\pm$0.00160 & 0.30230$\pm$0.07868 & 0.4657$\pm$0.0104 & 3102.71456$\pm$147.7188  & 2.02530  &  11  \\
    & 0.7 & 0.05841$\pm$0.00800  & 1.15780$\pm$0.00210 & 0.29330$\pm$0.01027& 0.4594$\pm$0.0102 & 3505.54287$\pm$743.5215  & 0.79604  &  11  \\
    & 0.9 & 0.05607$\pm$0.01276  & 1.16250$\pm$0.00210 & 0.28230$\pm$0.08254 & 0.4505$\pm$0.0098 & 4164.10151$\pm$498.1546  & 2.33216  &  11  \\
900 GeV    & 1.1 & 0.05413$\pm$0.01093  & 1.16750$\pm$0.00170 & 0.27330$\pm$0.06171 & 0.4452$\pm$0.0096 & 4582.32914$\pm$483.4055  & 1.86525  &  11  \\
    & 1.3 & 0.05291$\pm$0.01811  & 1.16820$\pm$0.00160 & 0.26430$\pm$0.07412 & 0.4363$\pm$0.0097 & 5265.66198$\pm$926.7622  & 5.55728  &  11  \\
    & 1.5 & 0.05099$\pm$0.01561  & 1.17397$\pm$0.00190 & 0.25540$\pm$0.04437 & 0.4326$\pm$0.0098 & 5929.58665$\pm$986.7167  & 4.46334  &  11  \\
    & 1.7  & 0.04787$\pm$0.02216  & 1.18117$\pm$0.00170 & 0.24440$\pm$0.06597 & 0.4241$\pm$0.0095 & 7377.07226$\pm$774.4555  & 10.74392  &  11  \\
    & 1.9 & 0.04630$\pm$0.01999  & 1.18512$\pm$0.00220 & 0.23140$\pm$0.04048 & 0.4178$\pm$0.0097 & 8426.60004$\pm$647.9168  & 9.20621  &  11  \\
    & 2.1 & 0.04593$\pm$0.02960  & 1.18614$\pm$0.00210 & 0.22680$\pm$0.02826 & 0.4159$\pm$0.0098 & 8631.31528$\pm$988.5175  & 20.35876  &  11  \\
    & 2.3  & 0.04462$\pm$0.02712  & 1.18842$\pm$0.00180 & 0.21980$\pm$0.07804 & 0.4103$\pm$0.0096 & 9324.19040$\pm$959.5391  & 17.08767  &  11  \\
\hline
\hline
    & 0.1  & 0.06808$\pm$0.01862  & 1.13570$\pm$0.00120 & 0.35320$\pm$0.06337 & 0.4988$\pm$0.0132 & 2521.38642$\pm$632.14545  & 11.70458  &  11  \\
    & 0.3  & 0.06666$\pm$0.01492  & 1.14120$\pm$0.00110 & 0.34320$\pm$0.03279 & 0.4962$\pm$0.0131 & 2737.22202$\pm$721.88793  & 11.15294  &  11  \\
    & 0.5 & 0.06376$\pm$0.01681  & 1.14920$\pm$0.00140 & 0.33143$\pm$0.05559 & 0.4894$\pm$0.0130 & 3153.96261$\pm$744.76038  & 9.25049  &  11  \\
    & 0.7    & 0.06191$\pm$0.01356  & 1.15390$\pm$0.00150 & 0.32320$\pm$0.05177 & 0.4841$\pm$0.0130 & 3573.01903$\pm$419.57793  & 5.62341  &  11  \\
    & 0.9 & 0.06064$\pm$0.01133  & 1.16190$\pm$0.00140 & 0.31520$\pm$0.04847 & 0.4880$\pm$0.0129 & 3823.19530$\pm$3021.7187  & 7.87698  &  11  \\  
    & 1.1  & 0.05850$\pm$0.01103  & 1.16690$\pm$0.00130 & 0.30420$\pm$0.03436 & 0.4806$\pm$0.0128 & 4205.86453$\pm$2683.5818  & 5.04605  &  11  \\   
 2.36 TeV   & 1.3  & 0.05579$\pm$0.02381  & 1.17230$\pm$0.00140 & 0.29280$\pm$0.03592 & 0.4707$\pm$0.0127 & 5075.72273$\pm$440.52248  & 5.98253  &  11  \\
    & 1.5& 0.05365$\pm$0.01198  & 1.17960$\pm$0.00130 & 0.28350$\pm$0.03698 & 0.4683$\pm$0.0126 & 5756.07442$\pm$505.87002  & 10.42228  &  11  \\
    & 1.7 & 0.05025$\pm$0.00855  & 1.18650$\pm$0.00120 & 0.27510$\pm$0.02376 & 0.4585$\pm$0.0125 & 7189.78232$\pm$466.20394  & 2.77283  &  11  \\
    & 1.9 & 0.04862$\pm$0.007163  & 1.19260$\pm$0.00120 & 0.26805$\pm$0.04759 & 0.4577$\pm$0.0123 & 7994.20741$\pm$426.04139  & 3.75365  &  11  \\
    & 2.1  & 0.04630$\pm$0.007202  & 1.19160$\pm$0.00130 & 0.25152$\pm$0.06712 & 0.4367$\pm$0.0122 & 9987.74237$\pm$881.8848  & 8.32422  &  11  \\
    & 2.3 & 0.04509$\pm$0.0052719  & 1.20030$\pm$0.00130 & 0.23545$\pm$0.06614 & 0.4399$\pm$0.0121 & 10601.8529$\pm$1992.9341  & 10.56820  &  11  \\
\hline
\hline
    & 0.1   & 0.06832$\pm$0.01981  & 1.14680$\pm$0.00160 & 0.35330$\pm$0.15715 & 0.5197$\pm$0.0141 & 3032.5352$\pm$1322.6312  & 35.64743  &  11  \\
    & 0.3  & 0.06677$\pm$0.01563  & 1.15380$\pm$0.00150 & 0.34330$\pm$0.15345 & 0.5192$\pm$0.0140 & 3309.0728$\pm$1337.3670  & 31.29893  &  11  \\
    & 0.5  & 0.06402$\pm$0.01590  & 1.16380$\pm$0.00170 & 0.33232$\pm$0.14413 & 0.5172$\pm$0.0139 & 3791.2659$\pm$1332.9927  & 25.72337  &  11  \\
    & 0.7    & 0.06217$\pm$0.01108  & 1.16680$\pm$0.00180 & 0.32430$\pm$0.13194 & 0.5091$\pm$0.0140 & 4232.3776$\pm$1321.5200  & 18.98844  &  11  \\
    & 0.9  & 0.06081$\pm$0.01061  & 1.17180$\pm$0.00160 & 0.31552$\pm$0.13338 & 0.5068$\pm$0.0138 & 4620.0929$\pm$1306.8054  & 14.87211  &  11  \\
    & 1.1  & 0.05867$\pm$0.010235  & 1.17780$\pm$0.00190 & 0.30456$\pm$0.13057 & 0.5013$\pm$0.0138 & 5143.6981$\pm$1318.5732  & 13.59588  &  11  \\   
 7 TeV   & 1.3  & 0.05662$\pm$0.01019  & 1.18120$\pm$0.00180 & 0.29345$\pm$0.12279 & 0.4916$\pm$0.0137 & 5963.8385$\pm$119.83735  & 10.29492  &  11  \\
    & 1.5 & 0.05441$\pm$0.01071  & 1.18720$\pm$0.00200 & 0.28435$\pm$0.12457 & 0.4866$\pm$0.0139 & 6822.2853$\pm$1415.1623  & 8.82797 & 11 \\
    & 1.7 & 0.05060$\pm$0.01011  & 1.19840$\pm$0.00220 & 0.27576$\pm$0.12001 & 0.4824$\pm$0.0138 & 8452.8398$\pm$1587.1667  & 7.00883  &  11  \\
    & 1.9 & 0.04948$\pm$0.01015  & 1.20140$\pm$0.00210 & 0.27016$\pm$0.11475 & 0.4795$\pm$0.0135 & 9092.0292$\pm$1398.6969  & 5.97471  &  11  \\
    & 2.1   & 0.04853$\pm$0.01062  & 1.20410$\pm$0.00210 & 0.25262$\pm$0.12012 & 0.4725$\pm$0.0134 & 9961.0765$\pm$1431.1808  & 6.80707  &  11  \\
    & 2.3 & 0.04584$\pm$0.01077  & 1.20860$\pm$0.00230 & 0.24830$\pm$0.12475 & 0.4640$\pm$0.0134 & 11652.4080$\pm$1642.1596  & 10.65240  &  11  \\    
\hline
\end{tabular}
\end{adjustbox}
\end{center}
\end{table*}

The results obtained from the fitting procedures, including the values of various free parameters, the normalization constant, standard deviation ($\chi^2$), and the number of degrees of freedom (ndof), have been given in Table 1. Specifically, Table 1 tabulates the data extracted from the fitting plots given in Fig. 1. The free parameters tabulated in Table 1 are the kinetic freeze-out temperature ($T_0$), transverse flow velocity ($\beta_T$) and non-extensivity parameter ($q$).

Fig. 2 gives the results obtained through fitting using the thermodynamically consistent modified Tsallis distribution function, corresponding to Fig. 1. This figure provides important dependencies of the extracted parameters with changing $\eta$. Fig. 1(a) represents $T_0$ as a function of $\eta$, where one can see that the earlier decreases with an increase in the latter. It is because of the maximum energy transfer into the system along the mid $\eta$ (low values) regions compared to forward or backwards (high values) $\eta$ regions. The same figure also represents an increasing trend of $T_0$ with increasing collision energy ($\sqrt{s}$). It is evident from the figure that the difference in $T_0$ is greater from $\sqrt{s}$ = 0.9 to 2.36 TeV but this difference in $T_0$ is minute as one moves from $\sqrt{s}$ = 2.36 to 7 TeV, i.e., the excitation function of $T_0$ is greater at lower energies compared to higher energies. This low excitation function of $T_0$ at higher energies may be attributed to the possible phase transition from hadronic to QCD (Quantum Chromodynamics) matter, where the extra energy is mostly used in the conversion of one phase of matter to another rather than appreciably increasing the temperature of the matter. 

Fig. 2(b) shows the decreasing trend of $\beta_T$ with increasing $\eta$, it is because, in the mid-rapidity regions the energy transfer is maximum which results in the production of a highly compressed system which expands rapidly, due to the huge pressure gradient, with greater $\beta_T$. The same figure also represents that the sensitivity of $\beta_T$ to $\sqrt{s}$ is greater at smaller $\sqrt{s}$ compared to larger $\sqrt{s}$. This may be a possible indication of the phase transition from hadrons to QGP at higher $\sqrt{s}$, where the extra energy is utilized for phase change rather than increasing the flow velocity of the system. 

Fig. 2(c) displays $q$ as a function of $\eta$ where the earlier increases with an increase in the latter. This means that the particles close to the mid $\eta$ are closer to thermal equilibrium compared to those which are along the beam direction. It is because, due to the greater energy transfer in the mid $\eta$ region, the produced particles have greater temperatures and quickly attain the temperature of their surrounding medium (QGP). On the other hand particles close to the beam direction have lower temperatures and take longer to establish equilibration with the surrounding medium. Moreover, Fig. 2(d) represents the fitting constant 
versus $\eta$. One can see that with increasing the latter, the earlier monotonically increases.

Fig. 2(e) represents the dependence of $\langle p_T \rangle$ on $\eta$. It is evident from the figure that with increasing the latter the earlier decreases due to the same reason as we discussed for $T_0$. The figure also shows that with increasing $\sqrt{s}$, $\langle p_T \rangle$ increases because of the greater number of multiple scatterings in the higher energy collisions.

Fig. 3 displays the correlations among the parameters extracted from the modified Tsallis distribution. Fig. 3(a) shows a direct correlation between $T_0$ and $\beta_T$, where one can see that with increasing the latter the earlier also increases. The values of $T_0$ and $\beta_T$ have a positive correlation. This is in line with the early universe when the system was rapidly expanding and sufficiently heated. Fig. 3(b) shows a negative correlation between $T_0$ and $q$. This renders that the parameters $T_0$ and $q$ in a system show stronger correlations between its components as it becomes more non-extensive ($q$ increases). The system's overall dynamics are impacted by the correlations, which lowers the temperature needed for freezeout. Where particles with smaller $q$ correspond to larger $T_0$ confirming that they are closer to equilibration and thermalization. Moreover, Fig. 3(c) is used to represent the negative correlation among different values of $\beta_T$ and $q$. The interaction of the QGP's expansion dynamics with the statistical characteristics of the particle distributions inside of it results in the negative correlation between the $\beta_T$ and the parameter $q$. Because of the departure from thermal equilibrium, the parameter $q$ rises as the system cools and expands while $\beta_T$ falls. This negative correlation can be explained by the fact that the parameter $q$ is influenced by the expansion dynamics, which also affect the distribution of particles in momentum space. On the other hand, the parameter $q$ may also affect the system's overall flow patterns. 
Fig. 3(d) displays the direct correlation between $T_0$ and $\langle p_T \rangle$ which means that the highly excited system always has greater $\langle p_T \rangle$. Fig. 3(e) represents the inverse correlation between $q$ and $\langle p_T \rangle$, providing evidence that the system closer to thermal equilibrium always has larger $\langle p_T \rangle$ i.e., highly excited. 
Fig. 3(f) shows the direct correlation between $\beta_T$ and $\langle p_T \rangle$, confirming the greater $\langle p_T \rangle$ for the rapidly expanding system, as expected.


\section{Conclusion}
\label{sec:conc}
In conclusion, our analysis of double differential \(p_T\) distributions in \(pp\) collisions at different center-of-mass energies reveals a strong agreement between experimental data and the modified Tsallis function. The strong agreement observed between experimental data and the modified Tsallis function, as the fit quality is consistently high across all \(p_T\) ranges, underscores the robustness of our approach. Through the fitting process, we have extracted essential parameters such as the kinetic freeze-out temperature (\(T_0\)), transverse flow velocity (\(\beta_T\)), non-extensivity parameter (\(q\)) and mean transverse momentum ($\langle p_T \rangle$), which provide valuable insights into the dynamics of the collision system. The findings also reveal significant dependencies of these parameters on pseudorapidity (\(\eta\)) and collision energy (\(\sqrt{s}\)), shedding light on the intricate interplay between kinematics and thermodynamics in high-energy collisions. \(T_0\), \(\beta_T\) and $\langle p_T \rangle$ decrease with increasing \(\eta\) due to smaller energy transfer along the beam direction. These parameters show an increasing trend with increasing \(\sqrt{s}\), suggesting the formation of a highly excited and compressed system at higher collision energies which at once releases all of its elastic potential energy in the transverse direction, resulting in the higher $\beta_T$. \(q\) increases with \(\eta\), indicating the particles produced along the mid-\(\eta\), away from the beam axis, to be closer to the thermal equilibrium. Correlation plots further highlight relationships between \(T_0\), \(\beta_T\), \(q\), and $\langle p_T \rangle$ emphasizing the complex dynamics of the system. These research findings contribute valuable insights into the thermal and dynamic characteristics of high-energy proton-proton collisions.

\section{Data availability}

All data analyzed during this study are included and/or properly cited in this article.

\section{Declaration of Interest Statement}
The authors declare that there are no conflicts of interest regarding the publication of this paper.

\section{Compliance with Ethical Standards}

The authors declare their adherence to ethical standards concerning the content of this paper.

\section{Acknowledgment}
This work is financially supported by Princess Nourah bint Abdulrahman University Researchers Supporting Project number (PNURSP2024R106), Princess Nourah bint Abdulrahman University, Riyadh, Saudi Arabia. 

\bibliographystyle{unsrt}
\bibliography{references}

\end{document}